# SARS-CoV-2 Entry Genes Are Most Highly Expressed in Nasal Goblet and Ciliated Cells within Human Airways

**Waradon Sungnak[1,†], Ni Huang[1], Christophe Bécavin[2], Marijn Berg[3,4], HCA Lung Biological Network*,[†]**

[1]Wellcome Sanger Institute, Wellcome Genome Campus, Hinxton, Cambridge, CB10 1SA, UK

[2]Université Côte d'Azur, CNRS, IPMC, Sophia-Antipolis 06560, France

[3]Department of Pathology and Medical Biology, University Medical Centre Groningen, University of Groningen, 9713 AV Groningen, Netherlands

[4]Groningen Research Institute for Asthma and COPD, University Medical Centre Groningen, University of Groningen, 9713 AV Groningen, Netherlands

[†]Authors for correspondence (ws4@sanger.ac.uk; lung@humancellatlas.org)

## Abstract

The SARS-CoV-2 coronavirus, the etiologic agent responsible for COVID-19 coronavirus disease, is a global threat. To better understand viral tropism, we assessed the RNA expression of the coronavirus receptor, *ACE2*, as well as the viral S protein priming protease *TMPRSS2* thought to govern viral entry in single-cell RNA-sequencing (scRNA-seq) datasets from healthy individuals generated by the Human Cell Atlas consortium. We found that *ACE2*, as well as the protease *TMPRSS2*, are differentially expressed in respiratory and gut epithelial cells. In-depth analysis of epithelial cells in the respiratory tree reveals that nasal epithelial cells, specifically goblet/secretory cells and ciliated cells, display the highest *ACE2* expression of all the epithelial cells analyzed. The skewed expression of viral receptors/entry-associated proteins towards the upper airway may be correlated with enhanced transmissivity. Finally, we showed that many of the top genes associated with *ACE2* airway epithelial expression are innate immune-associated, antiviral genes, highly enriched in the nasal epithelial cells. This association with immune pathways might have clinical implications for the course of infection and viral pathology, and highlights the specific significance of nasal epithelia in viral infection. Our findings underscore the importance of the availability of the Human Cell Atlas as a reference dataset. In this instance, analysis of the compendium of data points to a particularly relevant role for nasal goblet and ciliated cells as early viral targets and potential reservoirs of SARS-CoV-2 infection. This, in turn, serves as a biological framework for dissecting viral transmission and developing clinical strategies for prevention and therapy.

## Introduction

In December 2019, a cluster of atypical pneumonia associated with a novel coronavirus was detected in Wuhan, China[1]. This coronavirus disease, termed COVID-19, was caused by severe acute respiratory syndrome coronavirus 2 (SARS-CoV-2; previously termed 2019-nCoV)[2]. The virus has since spread worldwide, emerging as a serious global health concern in early 2020[3,4]. Human-to-human transmission of the virus has been reported in several instances[5-7] and is thought to have occurred since mid-December 2019[8]. As of early March 2020, there were more than 100,000 confirmed COVID-19 cases[4].

Patients with suspected COVID-19 have been treated in the Wuhan Jin Yintan Hospital since Dec 31st, 2019[9]. In a meta-analysis of 50,466 hospitalized patients with COVID-19 from 10 studies, most patients were from China and the average age in the included studies ranged from 41 to 56 years old[10]. The prevalence rates of fever, cough, and muscle soreness or fatigue were 89.1%, 72.2%, and 42.5%. Critical illness requiring admission to an intensive care unit occurred in 18.1% of patients, and 14.8% developed acute respiratory distress syndrome (ARDS)[10]. Acute renal injury and septic shock have been observed in 4% and 5% of patients hospitalized with COVID-19, respectively[1,9]. Chest imaging demonstrated bilateral pneumonia involvement in more than 80% of cases[1,9,11]. Ground-glass opacities were the most common radiologic finding on chest computed tomography (CT)[11,12]. Abnormalities on CT were also observed preceding symptom onset in patients exposed to infected individuals, with an incidence of 93%[10,11].

Pathological evaluation of a patient who died of severe disease revealed diffuse alveolar damage consistent with ARDS[13]. Currently, the estimated mortality rate is 3.4%[14]. These clinical data underscore the severity of this infection. The involvement of both lungs in most of the cases suggests viral dissemination after initial infection.

Viral RNA was detected in the upper airways from symptomatic patients, with higher viral loads observed in nasal swabs compared to those obtained from the throat[15]. Similar viral loads were observed in an asymptomatic patient[15], indicating that the nasal epithelium is an important portal for initial infection, and may serve as a key reservoir for viral spread across the respiratory mucosa and an important locus mediating viral transmission. Identification of the cells hosting viral entry and permitting viral replication as well as those contributing to inflammation and disease pathology is essential to improve diagnostic and therapeutic interventions.

Cellular entry of coronaviruses depends on the binding of the spike (S) protein to a specific cellular receptor and subsequent S protein priming by cellular proteases. Similar to severe acute respiratory syndrome-associated coronavirus (SARS-CoV)[16,17], the SARS-CoV-2 employs angiotensin-converting enzyme-2 (ACE2) as a receptor for cellular entry. In addition, studies have shown that the serine protease TMPRSS2 can prime S protein[15,18] although other proteases like cathepsin B/L can also be involved[18]. For SARS, the binding affinity between the S protein and the ACE2 receptor was found to be a major determinant of viral replication rates and disease severity[19]. The SARS-CoV-2 has been shown to infect and replicate in Vero cells, a *Cercopithecus aethiops* (old world monkey) kidney epithelial cell line, and huh7 cells, a human hepatocarcinoma cell line[15]. The BHK21 cell line has been shown to facilitate viral entry by the SARS-CoV-2 S protein only when engineered to express the ACE2 receptor ectopically[18]. In addition, viral entry was found to depend on TMPRSS2 activity, although cathepsin B/L activity might substitute for the loss of TMPRSS2[18]. The *in vivo* expression of ACE2 and TMPRSS2 (as well as other candidate proteases) by cells of the upper and lower airways and alveoli must be defined.

Previously, gene expression of *ACE2* and *TMPRSS2* has been reported to occur largely in type-2 alveolar (AT-2) epithelial cells[15], which are central to SARS-CoV pathogenesis. A study reported that ACE2 expression is absent from the upper airways[20]. The rapid spread of the SARS-CoV-2 suggests efficient human-to-human transmission which would, in turn, seem to supersede the odds of dependency on alveolar epithelial cells as the primary point of entry and viral replication[8,21,22]. Indeed, protein expression, based on immunohistochemistry, of ACE2 and TMPRSS2 has been reported in both nasal and bronchial epithelium[23]. To clarify the expression patterns of *ACE2* and *TMPRSS2* and analyze the expression of the other potential genes associated with SARS-CoV-2 pathogens at cellular resolution, we interrogated single-cell transcriptome expression data from published scRNA-seq datasets from healthy donors generated by the Human Cell Atlas consortium[24].

## Results

### *ACE2* and *TMPRSS2* are enriched in nasal tissues and enterocytes

We investigated the gene expression of *ACE2* in multiple scRNA-seq datasets from different tissues, including those of the respiratory tree[25], ileum[26], colon[27], liver[28], placenta/decidua[29], kidney[30], testis[31], pancreas[32], and prostate gland[33]. While scRNA-seq is a comprehensive assay, we note that some studies may still miss specific cell types, due to either their rarity, challenges associated with their isolation, or analysis methodology that was used. Thus, while positive (presence) results are highly reliable, absence should be interpreted with care.

The expression of *ACE2*, in general, is relatively low in all of the datasets analyzed. Consistent with independent analyses[34], we found that *ACE2* is expressed in lung, airways, ileum, colon, and kidney (**Fig. 1a**; first column). It is worth noting that *TMPRSS2*, the primary protease important for viral entry, is highly expressed with a broader distribution (**Fig. 1a**; second column), suggesting that *ACE2*, rather than *TMPRSS2*, may be a limiting factor for viral entry at the initial stage of infection. When taking into account the expression of both genes, the cells found in mucosal epithelia in the respiratory tree, ileum, and colon are *ACE2*⁺ (**Fig. 1a**; third column), consistent with viral transmission by respiratory droplets, and the potential of fecal-oral transmission[35]. We also assessed *ACE2* and *TMPRSS2* expression in developmental datasets from fetal liver, fetal thymus, fetal skin, fetal bone marrow and fetal yolk sac[36,37] and found little to no expression of *ACE2* with no co-expression with *TMPRSS2* (data not shown) even if single *ACE2* expression is noticeable in certain cell types in placenta/decidua (**Fig. 1a**). While we cannot rule out the possibility that the virus uses alternative proteases for entry in such contexts, or that lung fetal tissue expresses the relevant genes, these results are at least consistent with early reports that fail to detect evidence of intrauterine infection through vertical transmission in women who develop COVID-19 pneumonia in late pregnancy[38]. If future epidemiologic data are consistent with a lack vertical viral transmission, these findings may form the basis of an explanatory model for the clinical finding. However, if future evidence for vertical transmission emerges, additional scRNA-seq data can be collected and further scrutinized for the presence of rare co-expressers or alternative receptors or proteases.

### Nasal goblet and ciliated cells display the highest expression of *ACE2* within the larger population of respiratory epithelial cells

To further characterize specific epithelial cell types expressing *ACE2*, we evaluated the expression of *ACE2* within lung/airway epithelia from a previous study[25]. We found that, despite a low level of expression overall, *ACE2* is expressed in multiple epithelial cell types across the airway, as well as in AT-2 cells in the parenchyma, consistent with previous studies[20,39]. Importantly, nasal epithelial cells, including previously described two clusters of goblet cells and one cluster of ciliated cells, have the highest expression among all investigated cells in the respiratory tree (**Fig. 1b**; left panel). We confirmed enriched *ACE2* expression in nasal epithelial cells from a second scRNA-seq study, which, in addition to nasal brushing samples seen in the earlier dataset, included nasal biopsies[40]. The results were consistent: we found the highest expression of *ACE2* in nasal secretory cells (equivalent to the two goblet cell clusters in the previous dataset) and ciliated cells (**Fig. 1b**; right panel).

In addition, scRNA-seq data from an *in vitro* 3D epithelial regeneration system from nasal epithelial cells[41] corroborated the expression of *ACE2* in goblet/secretory cells and ciliated cells in these air-liquid interface (ALI) cultures (**Extended Data Fig. 1**). Of note, the differentiating cells in ALI acquire progressively more *ACE2* and, unlike their corresponding progenitors, they have large luminal surfaces in the mature differentiated epithelium where viral entry is likely to occur (**Extended Data Fig. 1**). These results also suggest that such *in vitro* culture system is biologically relevant to the study of viral pathogenesis.

We also investigated the expression of known proteases associated with the entry of SARS-CoV and SARS-CoV-2. *TMPRSS2*, which was shown to be important for SARS-CoV/SARS-CoV-2 viral entry and SARS-CoV transmission,[16-18] is expressed in a subset of *ACE2*$^{+}$ cells (**Extended Data Fig. 2**), suggesting that the virus might use alternative pathways for entry. In fact, it was previously shown that SARS-CoV-2 could enter TMPRSS2$^{-}$ cells using cathepsin B/L[18]. Indeed, we found that they are much more promiscuously expressed than *TMPRSS2*, especially cathepsin B, which is expressed in more than 70%-90% of *ACE2*$^{+}$ cells (**Extended Data Fig. 2**). However, whether cathepsin B/L can functionally replace TMPRSS2 has not been empirically determined. In the case of SARS-CoV, TMPRSS2 activity is documented to be important for viral transmission[42,43].

**Respiratory expression of viral receptor/entry-associated genes and implications for viral transmissivity**

We next asked whether the enriched expression of viral receptors and entry-associated molecules in the nasal region/upper airway could be relevant to viral transmissivity. Here, we assessed the expression of viral receptor genes that are used by other coronaviruses and influenza viruses, including *ANPEP* (used by HCoV-229[44]) and *DPP4* (used by MERS-CoV[45]), as well as the enzymes *ST6GAL1* and *ST3GAL4* in the lung epithelial cell datasets. The latter genes are enzymes which are important for the synthesis of viral receptors used by influenza viruses: α(2,6)-linked sialic acid and α(2,3)-linked sialic acid[46]. Notably, the distribution of receptor/receptor-associated enzymes appears to coincide with viral transmissivity patterns based on a comparison to the basic reproduction number ($R_0$), which estimates the number of people who can get infected from a single infected person; and the infection will be able to start spreading in a population when $R_0 > 1$. The skewed distribution of the receptors/enzymes towards the upper airway is observed in viruses with relatively higher $R_0$/infectivity, including those of SARS-CoV/SARS-CoV-2 ($R_0$ ~ 1.4-5.0[8,21,22]), influenza (mean $R_0$ ~1.3[47]) and HCoV-229E (unidentified $R_0$; associated with common cold[48]). This distribution is in distinct contrast with that of *DPP4*, the receptor for MERS-CoV ($R_0$ ~0.3-0.8), a coronavirus with limited human-to-human transmission[49], with the skewed expression towards lower airway/lung parenchyma (**Fig. 2a**). Therefore, our data highlight the possibility that viral transmissivity is dependent on receptor accessibility based on spatial distribution along the respiratory tract.

## Expression of genes associated with *ACE2* expression: innate immunity and carbohydrate metabolism

To gain more insight into the expression patterns of genes associated with *ACE2*, we performed Spearman correlation analysis with Benjamini-Hochberg-adjusted *p*-values on genes associated with *ACE2* across all cells within the lung epithelial cell dataset[25]. While the correlation coefficients are relatively low (< 0.11), likely due to low expression of ACE2, the expression pattern of the top 50 *ACE2*-correlated genes (all with adjusted *p*-value close to 0; ranked by correlation coefficients) across the respiratory tree is similar to that of *ACE2*, with a skewed expression toward upper airway (**Fig. 2b**). To our surprise, while some of the genes are associated with carbohydrate metabolism (possibly due to the role of goblet cells in mucin synthesis), a number of genes associated with immune functions including innate and antiviral immune functions, are over-represented in the rank list, including *IDO1*, *IRAK3*, *NOS2*, *TNFSF10*, *OAS1*, and *MX1* (**Fig. 2b** and **Supplementary Table 1**). These genes have the highest expression in nasal goblet 2 cells (**Fig. 2b**), consistent with the phenotype previously described[25]. Nonetheless, nasal goblet 1 and nasal ciliated 2 cells also significantly express these genes, but less so elsewhere (**Fig. 2b**). Given their environmental exposure and the high expression of receptor/receptor-associated enzymes (**Fig. 2a**), it is plausible that the nasal epithelial cells were conditioned and primed to express these immune-associated genes to prevent viral susceptibility. This association with innate immune pathways not only highlights the importance of host-microbe dynamics in nasal epithelia, but it may also have implications for subsequent viral pathogenesis and immune-associated protection/pathology.

## Discussion

In this study, we explored multiple scRNA-seq datasets generated within the HCA consortium, and found that SARS-CoV-2 entry receptor *ACE2* is more highly expressed (and co-expressed with viral entry-associated protease *TMPRSS2)* in nasal epithelial cells, specifically goblet and ciliated cells. This finding implicates these cells as loci of original infection and possible reservoirs for dissemination within a given patient and from person to person. Importantly, viral infection itself could drastically change the gene expression landscape in the nose and other tissues later on.

The up-regulation of innate immune genes, in association with *ACE2*, in highly-exposed nasal epithelial cells could be the result of their responsiveness to persistent environmental challenges, including viral infection. It would be of great interest to further investigate how other genetic, demographic, and environmental factors might affect this poised state in these cells and whether such state could influence the susceptibility to infection due to its association with viral receptor expression. Future meta-analysis of HCA data can help further assess some of these aspects.

All in all, our findings may have significant implications for understanding viral transmissivity, considering that the primary viral transmission is through respiratory droplets. Moreover, as SARS-CoV-2 is an enveloped virus, its release does not require cell lysis. Thus, the virus might exploit existing secretory pathways in nasal goblet cells for low-level, continuous-release at the early stage with no overt pathology. These discoveries could have clinical implications with respect to targeting nasal epithelial cells, especially nasal goblet cells, beyond the current usage of face masks, providing a candidate clinical option for transmission prevention and/or early-stage intervention.

Finally, it is worth highlighting that this is the first collaborative effort by a Human Cell Atlas Biological Network (the Lung), and illustrates the opportunities from integrative analyses of Human Cell Atlas data, with future examples of consortium work expected soon.

**Methods**

The datasets were retrieved from existing sources based on previously published data as specifically specified in the reference. We retained the cell clustering when available or reprocessed using scanpy[50] and harmony[51], and annotated the clusters with marker genes and cell type nomenclature based on the respective studies. Illustration of the results was generated using scanpy[50] and Seurat[52].

**Acknowledgements**

We are grateful to Cori Bargmann, Jeremy Farrar, and Sarah Aldridge for stimulating discussions. We thank Jana Eliasova (scientific illustrator) for support with the design of figures.

This publication is part of the Human Cell Atlas - www.humancellatlas.org/publications.

*The HCA Lung Biological Network are: Pascal Barbry[1], Alvis Brazma[2], Tushar Desai[3], Thu Elizabeth Duong[4], Oliver Eickelberg[5], Muzlifah Haniffa[6], Peter Horvath[7], Naftali Kaminski[8], Mark Krasnow[9], Malte Kuhnemund[10], Haeock Lee[11], Sylvie Leroy[12], Joakim Lundeberg[13], Kerstin B. Meyer[14], Alexander J. Misharin[15], Martijn C. Nawijn[16], Marko Z. Nikolic[17], Jose Ordovas Montanes[18], Dana Pe'er[19], Joseph Powell[20], Steve Quake[21], Jay Rajagopal[22], Purushothama Rao Tata[23], Emma L. Rawlins[24], Aviv Regev[25], Orit Rozenblatt-Rosen[26], Kourosh Saeb-Parsy[27], Christos Samakovlis[28], Herbert B. Schiller[29], Joachim L. Schultze[30], Alex K. Shalek[31], Douglas Shepherd[32], Xin Sun[33], Sarah A. Teichmann[34], Fabian Theis[35], Alexander Tsankov[36], Maarten van den Berge[37], Jeffrey Whitsett[38], and Kun Zhang[39].

## Affiliations

[1]Université Côte d'Azur, CNRS, IPMC, Sophia-Antipolis 06560, France
[2]European Molecular Biology Laboratory, European Bioinformatics Institute (EMBL-EBI), Wellcome Trust Genome Campus, Hinxton, Cambridge, CB10 1SD, UK

[3]Department of Medicine and Institute for Stem Cell Biology and Regenerative Medicine, Stanford University School of Medicine, Stanford, CA 94116, USA
[4]Department of Pediatrics Division of Respiratory Medicine, University of California San Diego and Rady Children's Hospital San Diego, San Diego, CA 92123, USA
[5]Division of Pulmonary Sciences and Critical Care Medicine, Department of Medicine, University of Colorado, Anschutz Medical Campus, Aurora, CO, US
[6]Wellcome Sanger Institute, Wellcome Genome Campus, Hinxton, Cambridge CB10 1SA, UK; Biosciences Institute, Faculty of Medical Sciences, Newcastle University, Newcastle upon Tyne NE2 4HH, UK; Department of Dermatology and NIHR Newcastle Biomedical Research Centre, Newcastle Hospitals NHS Foundation Trust, Newcastle upon Tyne NE2 4LP, UK
[7]Synthetic and Systems Biology Unit, Biological Research Centre (BRC), Szeged, Hungary; Institute for Molecular Medicine Finland, University of Helsinki
[8]Pulmonary, Critical Care and Sleep Medicine, Yale University School of Medicine, New Haven, CT 06520, USA
[9]Department of Biochemistry and Wall Center for Pulmonary Vascular Disease, Howard Hughes Medical Institute, Stanford University School of Medicine, Stanford, CA 94305, USA
[10]Cartana AB, Nobels vag 16, 17165 Stockholm, Sweden
[11]Department of Biomedicine and Health Sciences, The Catholic University of Korea, Seoul, Korea
[12]Université Côte d'Azur, CHU de Nice, FHU OncoAge, Department of Pulmonary Medicine and Allergology, Nice, France; CNRS UMR 7275 - Institut de Pharmacologie Moléculaire et Cellulaire, Sophia Antipolis, France
[13]SciLifeLab, Department of Gene Technology, KTH Royal Institute of Technology, SE-100 44, Stockholm, Sweden
[14]Wellcome Sanger Institute, Wellcome Genome Campus, Hinxton, Cambridge, CB10 1SA, UK
[15]Division of Pulmonary and Critical Care Medicine, Northwestern University, Chicago, Illinois 60611, USA
[16]Department of Pathology and Medical Biology, University of Groningen, GRIAC Research Institute, University Medical Center Groningen, 9713 AV Groningen, Netherlands
[17]UCL Respiratory, Division of Medicine, University College London, WC1E 6JF, London, UK
[18]Division of Gastroenterology Boston Children's Hospital, Boston, MA 02115, USA; Program in Immunology, Harvard Medical School, Boston, MA 02115, USA; Broad Institute of MIT and Harvard, Cambridge, MA 02142, USA; Harvard Stem Cell Institute, Cambridge, MA 02138, USA.

[19]Computational and Systems Biology Program, Sloan Kettering Institute, Memorial Sloan Kettering Cancer Center, New York, New York 10065, USA
[20]Garvan-Weizmann Centre for Cellular Genomics, Garvan Institute of Medical Research, Sydney, NSW, Australia; UNSW Cellular Genomics Futures Institute, University of New South Wales, Sydney, NSW, Australia
[21]Depts of Bioengineering and Applied Physics, Stanford University, and the Chan Zuckerberg Biohub, Stanford University, Stanford, CA 94305, USA
[22]Harvard Stem Cell Institute, Cambridge, MA 02138, USA; Center for Regenerative Medicine, Massachusetts General Hospital, Boston, MA 02114, Boston
[23]Department of Cell Biology, Regeneration Next Initiative, Duke University School of Medicine, Durham, NC 27710, USA
[24]Wellcome Trust/CRUK Gurdon Institute and Department Physiology, Development and Neuroscience, University of Cambridge, Cambridge, CB2 1QN, UK
[25]Klarman Cell Observatory, Broad Institute of MIT and Harvard, Howard Hughes Medical Institute, Department of Biology, Massachusetts Institute of Technology, Cambridge, MA 02142, USA
[26]Klarman Cell Observatory, Broad Institute of Harvard and MIT, Cambridge, MA 02142, USA
[27]Department of Surgery, University of Cambridge and NIHR Cambridge Biomedical Research Centre, CB2 0QQ, UK
[28]SciLifeLab, Department of Molecular Biosciences, Stockholm University, Stockholm Sweden; Cardiopulmonary Institute, Justus Liebig University, Giessen, Germany
[29]Comprehensive Pneumology Center (CPC) / Institute of Lung Biology and Disease (ILBD), Helmholtz Zentrum München, Member of the German Center for Lung Research (DZL), Munich, Germany
[30]Joachim L. Schultze, 1 Department for Genomics & Immunoregulation, LIMES-Institute, University of Bonn, 53115 Bonn, Germany; 2 PRECISE Platform for Single Cell Genomics & Epigenomics, German Center for Neurodegenerative Diseases and University of Bonn, Bonn, Germany
[31]Ragon Institute of MGH, MIT, and Harvard, Cambridge, MA, USA; Institute for Medical Engineering and Science (IMES), Koch Institute for Integrative Cancer Research, and Department of Chemistry, Massachusetts Institute of Technology, Cambridge, MA, USA; Broad Institute of MIT and Harvard, Cambridge, MA, USA
[32]Center for Biological Physics and Department of Physics, Arizona State University, Tempe, AZ 85287, USA
[33]Department of Pediatrics, Department of Biological Sciences, University of California SD, 9500 Gilman Dr. MC0766, San Diego, CA 92093, USA

[34]Wellcome Sanger Institute, Wellcome Genome Campus, Hinxton, Cambridge, CB10 1SA, UK; Theory of Condensed Matter Group, Cavendish Laboratory/Department of Physics, University of Cambridge, Cambridge CB3 0HE, UK
[35]Institute of Computational Biology, Helmholtz Zentrum München and Departments of Mathematics and Life Sciences, Technical University Munich, Germany
[36]Genetics and Genomic Sciences, Icahn School of Medicine at Mount Sinai, New York, NY 10029, USA
[37]Department of Pulmonary diseases and tuberculosis, University of Groningen, GRIAC Research Institute, University Medical Center Groningen, 9713 AV Groningen, Netherlands
[38]Cincinnati Children's Hospital Medical Center, Cincinnati, OH 45229, USA
[39]UCSD Department of Bioengineering, 9500 Gilman Drive, MC0412, PFBH402, La Jolla, CA 92093, USA

Pascal Barbry, Alexander Misharin, Martijn Nawijn and Jay Rajagopal serve as the coordinators for the HCA Lung Biological Network.

## Funding

This work was supported by a Seed Network grant from the Chan Zuckerberg Initiative to P.B., T.D., T.E.D., O.E., P.H., N.K., M.K., K.B.M., A.M., M.C.N., D.P., J.R., P.R.T., S.Q., A.R., O.R., H.B.S., D.S., A.T., J.W. and K.Z. and by the European Union's H2020 research and innovation programme under grant agreement No 874656 (discovAIR) to P.B., A.B., M.K., S.L., J.L., K.B.M., M.C.N., K.S.P., C.S., H.B.S., J.S., F.T. and M.vd.B. W.S. acknowledges funding from the Newton Fund, Medical Research Council (MRC), The Thailand Research Fund (TRF), and Thailand's National Science and Technology Development Agency (NSTDA). M.C.N acknowledge funding from GSK Ltd, Netherlands Lung Foundation project no. 5.1.14.020 and 4.1.18.226. T.D. acknowledges funding from HubMap consortium and Stanford Child Health Research Institute- Woods Family Faculty Scholarship. T.E.D. acknowledges funding from HubMap. P.H. acknowledges funding from LENDULET-BIOMAG Grant (2018-342) and the European Regional Development Funds (GINOP-2.3.2-15-2016-00006, GINOP-2.3.2-15-2016-00026, GINOP-2.3.2-15-2016-00037). N.K. acknowledges funding from NIH grants R01HL127349, U01HL145567 and an unrestricted grant from Three Lakes Foundation. M.K. acknowledges HHMI and Wall Center for Pulmonary Vascular Disease. H.L.

acknowledges funding from National Research Foundation of Korea. K.M. acknowledges funding from Wellcome Trust. A.M. acknowledges funding from NIH grants HL135124, AG049665 and AI135964. M.Z.N. acknowledges funding from Rutherford Fund Fellowship allocated by the Medical Research Council and the UK Regenerative Medicine Platform (MR/ 5005579/1 to M.Z.N.). J.O.-M. acknowledges funding from Richard and Susan Smith Family Foundation. D.P. acknowledges funding from Alan and Sandra Gerry Metastasis and Tumor Ecosystems Center. J.P. acknowledges funding from National Health and Medical Research Council. P.R.T. acknowledges funding from R01HL146557 from NHLBI/NIH. E.L.R. acknowledges funding from MRC MR/P009581/1 and MR/SO35907/1. A.R. and O. R. acknowledge HHMI, the Klarman Cell Observatory, and the Manton Foundation. K.S.-P. acknowledges NIHR Cambridge Biomedical Research Centre. C.S. acknowledges Swedish research Council, Swedish Cancer Society, and CPI. H.B.S. acknowledges German Center for Lung Research and Helmholtz Association. J.S. acknowledges Boehringer Ingelheim, by the German Research Foundation (DFG; EXC2151/1, ImmunoSensation2 - the immune sensory system, project number 390873048), project numbers 329123747, 347286815) and by the HGF grant sparse2big. A.K.S. acknowledges the Beckman Young Investigator Program, a Sloan Fellowship in Chemistry, the NIH (5U24AI118672), and the Bill and Melinda Gates Foundation. F.T. Theis the German Center for Lung Research. M.vd.B. acknowledges from Ministry of Economic Affairs and Climate Policy by means of the PPP. J.W. acknowledges NIH, U01 HL148856 LungMap Phase II.

## Competing interests

N.K. was a consultant to Biogen Idec, Boehringer Ingelheim, Third Rock, Pliant, Samumed, NuMedii, Indaloo, Theravance, LifeMax, Three Lake Partners, Optikira in the last three years and received non-financial support from MiRagen. J.L. is a scientific consultant for 10X Genomics Inc. A.R. is a co-founder and equity holder of Celsius Therapeutics, an equity holder in Immunitas, and an SAB member of ThermoFisher Scientific, Syros Pharmaceuticals, Asimov, and Neogene Therapeutics. O.R. is a co-inventor on patent applications filed by the Broad Institute to inventions relating to single cell genomics applications, such as in PCT/US2018/060860 and US Provisional Application No. 62/745,259. A.K.S. reports compensation for consulting and/or SAB membership from Merck, Honeycomb Biotechnologies, Cellarity, Cogen Therapeutics, Orche Bio, and Dahlia Biosciences. F.T. reports receiving consulting fees from Roche Diagnostics GmbH, and ownership interest in Cellarity Inc. S.A.T. was a consultant at Genentech, Biogen and Roche in the last three years.

## Author Contributions

W.S., N.H., C.B., and M.B. performed data analyses. W.S, N.H. and the HCA Lung Biological Network interpreted the data. W.S., with significant input from the HCA Lung Biological Network, wrote the paper. All authors read the manuscript, offered feedback, and approved it before submission.

**Figure 1**

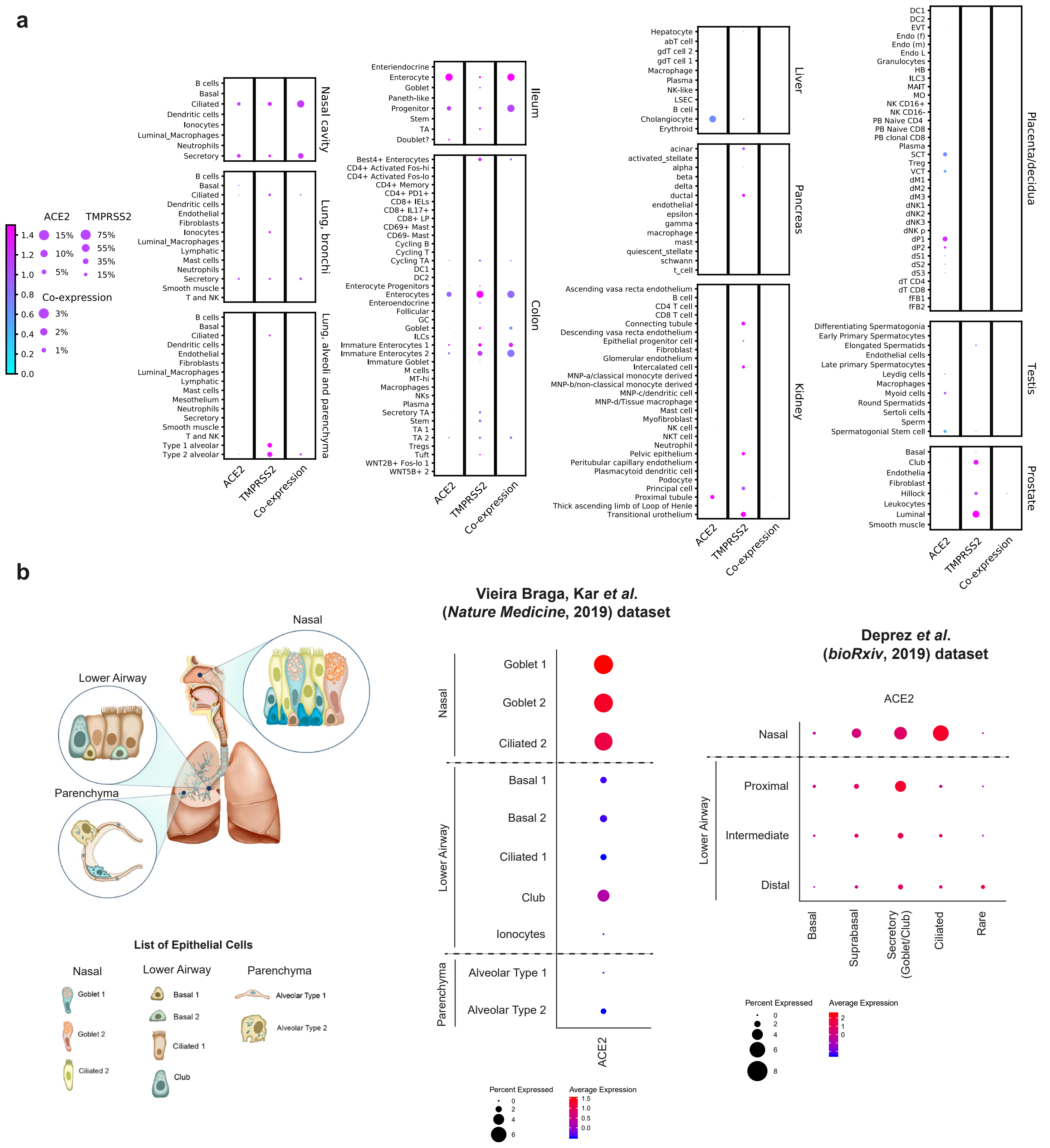

**Fig. 1| Expression of *ACE2* and *TMPRSS2* across different tissues and its enrichment in nasal epithelial cells. a**, RNA expression of SARS-CoV-2 entry receptor *ACE2* (first column), entry-associated protease *TMPRSS2* (second column), and their co-expression (third column) from multiple published scRNA-seq datasets. Raw expression values were normalized, log transformed and summarized by published cell clustering where available, or reproduced clustering annotated using marker genes and cell type nomenclature from the respective studies. The size of the dots indicates the proportion of cells in the respective cell type having greater-than-zero expression of *ACE2* (first column), *TMPRSS2* (second column) or both (third column), while the colour indicates the mean expression of ACE2 (first and third columns) or *TMPRSS2* (second column). **b**, Schematic illustration depicts the major anatomical regions in the human respiratory tree demonstrated in this study: nasal, lower airway, and lung parenchyma (left panel). Expression of *ACE2* is from airway epithelial cell datasets: Vieira Braga, Kar *et al.* 2019 (middle panel) and Deprez *et al.* 2019 (right panel). The datasets were retrieved from existing sources, and the cell clustering and nomenclature were retained based on the respective studies. For gene expression results in the dot plots: the dot size represents the proportion of cells within the respective cell type expressing the gene and the dot color represents the average gene expression level within the particular cell type.

Figure 2

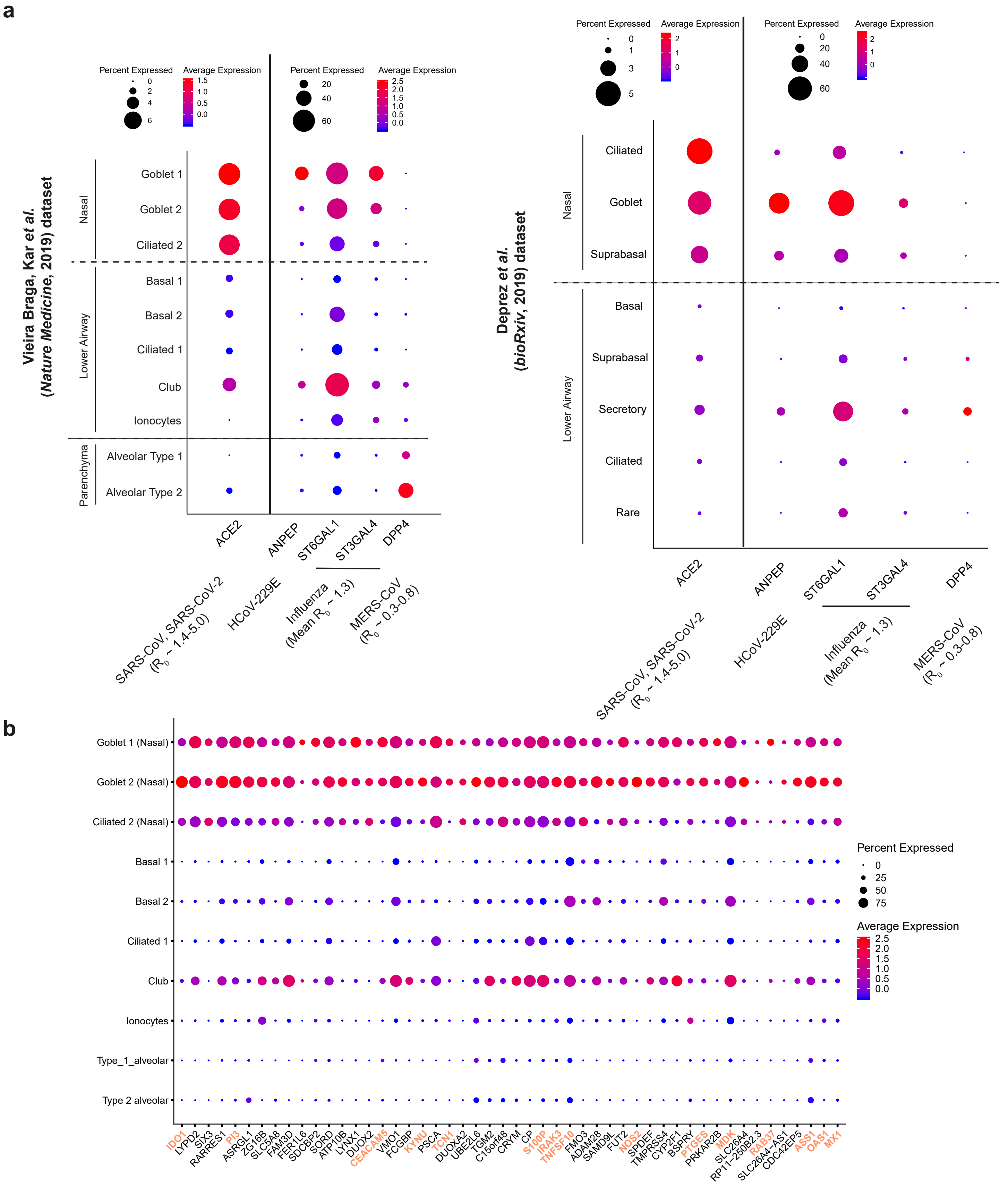

**Fig. 2| Respiratory expression of viral receptor/entry-associated genes and implications for viral transmissivity and genes associated with *ACE2* expression. a**, Expression of *ACE2* (an entry receptor for SARS-CoV and SARS-CoV-2), *ANPEP* (an entry receptor for HCoV-229E), *ST6GAL1*/*ST3GAL4* (enzymes important for synthesis of influenza entry receptors), and *DPP4* (an entry receptor for MERS-CoV) from the airway epithelial datasets: Vieira Braga, Kar *et al.* 2019 (left panel) and Deprez *et al.* 2019 (right panel). The basic reproductive number ($R_0$) for respective viruses, if available, are shown. **b**, Respiratory epithelial expression of the top 50 genes correlated with *ACE2* expression based on Spearman correlation analysis (with Benjamini-Hochberg-adjusted *p*-values) on genes associated with *ACE2* across all cells within the Vieira Braga, Kar *et al.* lung epithelial dataset. The colored gene names represent genes that are immune-associated (GO:0002376: immune system process or GO:0002526: acute inflammatory response). For gene expression results in the dot plots: the dot size represents the proportion of cells within the respective cell type expressing the gene and the color represents the average gene expression level within the particular cell type.

## Extended Data Figure 1

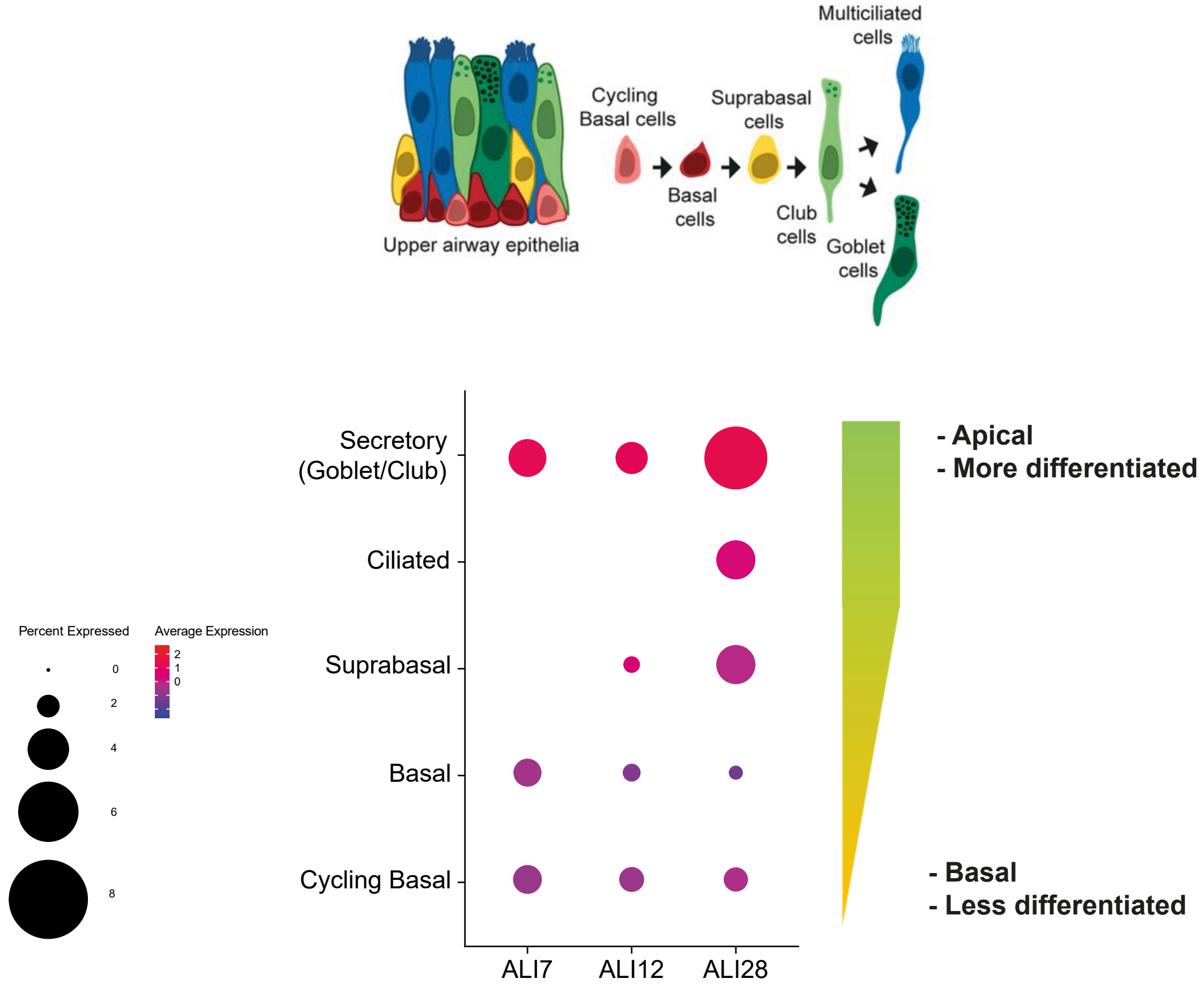

**Extend Data Fig. 1| Gene expression of *ACE2* in an *in vitro* 3D air-liquid interface (ALI) system.** Epithelial regeneration system from nasal epithelial cells was used for *in vitro* cultures on successive days (7, 12 and 28), resulting in different epithelial cell types along differentiation trajectory characterized in Ruiz García *et al.* 2019. The cultures were differentiated in Pneumacult media. Schematic illustration depicts the respective cell types in the differentiation trajectory, and the dot plot illustrates the cultured cell types along the differentiation pseudotime, along with their respective location within the epithelial layers. For gene expression results in the dot plot: the dot size represents the proportion of cells within the respective cell type expressing the gene and the dot color represents the average gene expression level within the particular cell type.

# Extended Data Figure 2

**Vieira Braga, Kar *et al.* (*Nature Medicine*, 2019) dataset**

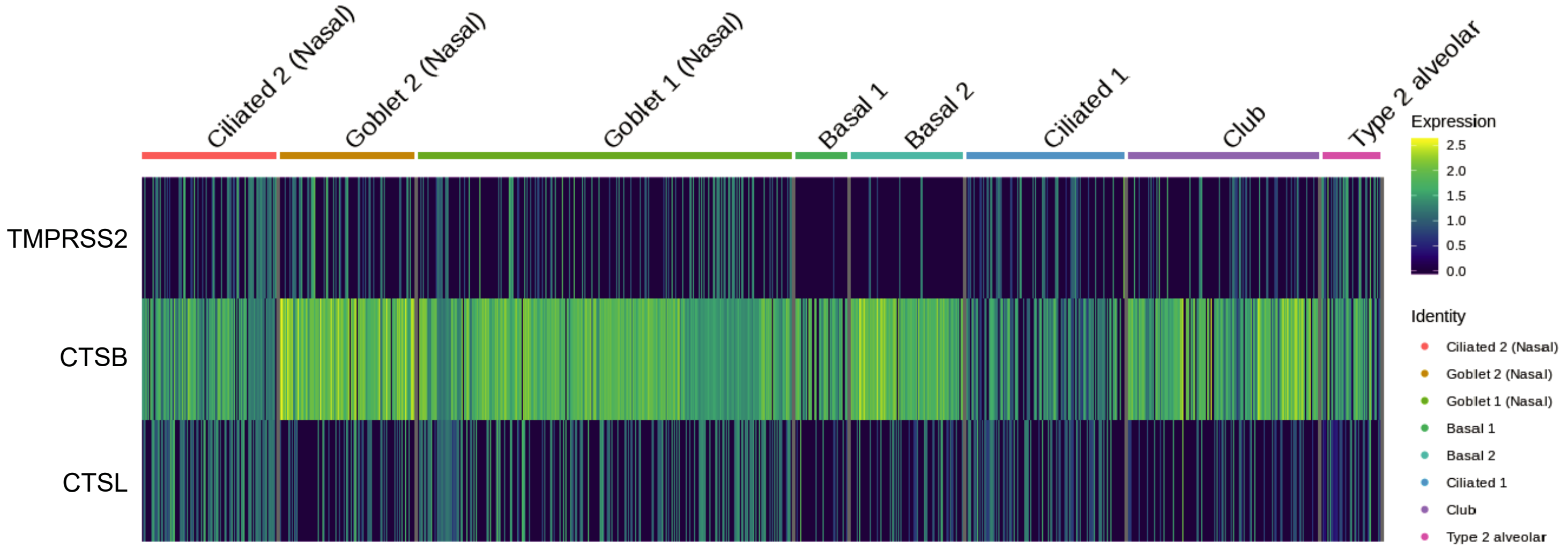

**Deprez *et al.* (*bioRxiv*, 2019) dataset**

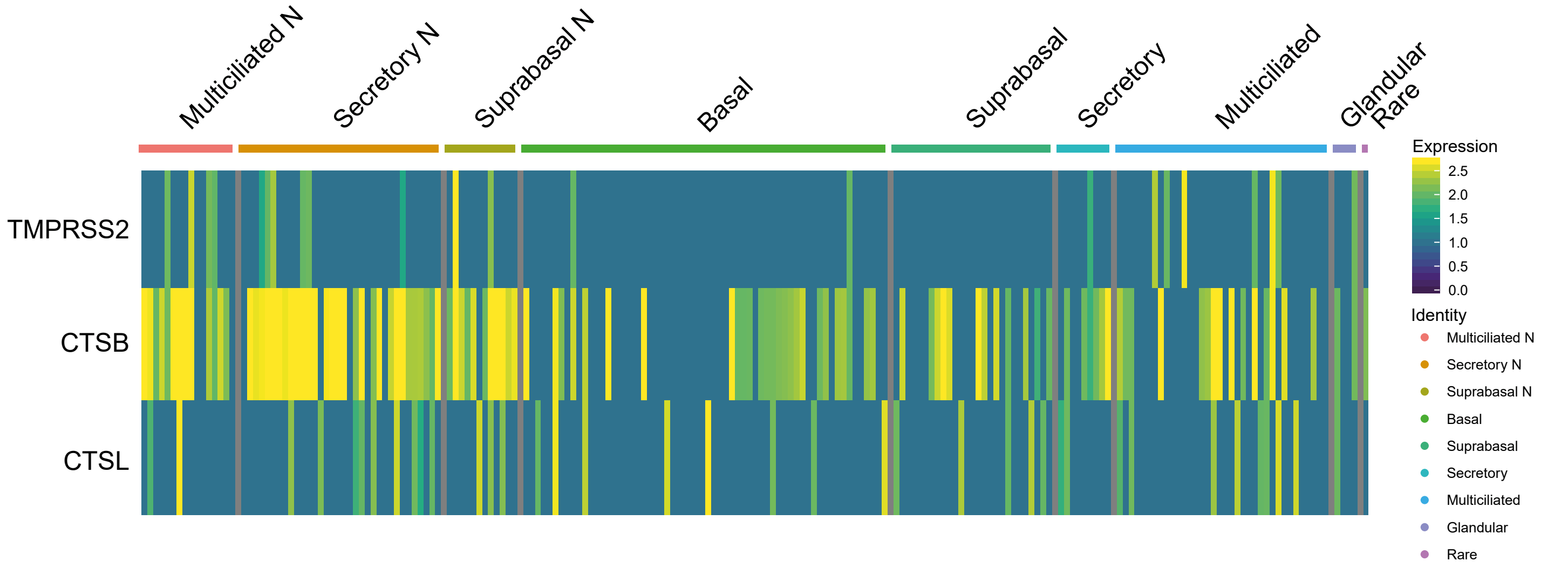

**Extended Data Fig. 2|** Expression and co-expression of SARS-CoV-2 entry-associated proteases in *ACE2*$^+$ airway epithelial cells: *TMPRSS2*, *CTSB*, and *CTSL* in *ACE2*$^+$ cells from the Vieira Braga, Kar *et al.* (top) and Deprez *et al.* (bottom) airway epithelial datasets. The color represents the expression level at the single-cell resolution and the cells are grouped based on the cell types specified.

## Supplementary Table 1

| Genes | GO Accession Number: Class | PathCards |
|---|---|---|
| IDO1 | GO:0002376: immune system process | NF-kappaB Signaling<br>Viral mRNA translation |
| PI3 | GO:0002376: immune system process | Defensins, Innate Immune System |
| CEACAM5 | GO:0002376: immune system process | NF-kappaB Signaling |
| KYNU | GO:0002376: immune system process | Viral mRNA translation |
| TCN1 | GO:0002376: immune system process | Innate Immune System<br>NF-kappaB Signaling |
| S100P | GO:0002376: immune system process | Innate Immune System |
| IRAK3 | GO:0002376: immune system process | Innate Immune System |
| TNFSF10 | GO:0002376: immune system process | TNF signaling |
| NOS2 | GO:0002376: immune system process | Innate Immune System |
| PTGES | GO:0002526: acute inflammatory response | Prostaglandin 2 biosynthesis and metabolism FM |
| MDK | GO:0002376: immune system process | NF-KappaB Family Pathway |
| RAB37 | GO:0002376: immune system process | Innate Immune System |
| ASS1 | GO:0002376: immune system process | Viral mRNA Translation |
| OAS1 | GO:0002376: immune system process | Innate Immune System<br>Immune response IFN alpha/beta signaling pathway |
| MX1 | GO:0002376: immune system process | Innate Immune System<br>Immune response IFN alpha/beta signaling pathway |

**Supplementary Table 1|** Immune-associated genes in respiratory epithelial expression from the top 50 genes correlated with *ACE2* expression based on Spearman correlation analysis (with Benjamini-Hochberg-adjusted *p*-values) across all cells within the Vieira Braga, Kar *et al.* lung epithelial dataset. The characterization of genes was based on Gene Ontology classes from the Gene Ontology (GO) database and associated pathways in PathCards from the Pathway Unification Database.